\documentclass[%
 reprint,
 amsmath,amssymb,
 aps,
floatfix,
]{revtex4-1}

\usepackage{graphicx}
\usepackage{dcolumn}
\usepackage{bm}
\usepackage{float}

\usepackage{natbib}

\begin{document}

\preprint{APS/123-QED}

\title{Morphology and kinetics of asphalt binder microstructure at gas, liquid, and solid interfaces}

\author{A. Ramm}
\author{M. C. Downer}%
 \email{downer@physics.utexas.edu}
\affiliation{%
Department of Physics,\\
 The University of Texas at Austin
}
\author{N. Sakib}

\author{A. Bhasin}
\affiliation{%
Department of Civil Engineering,\\
 The University of Texas at Austin
}




\date{\today}

\begin{abstract}

We combined optical and atomic force microscopy to observe morphology and kinetics of microstructures that formed at free surfaces of unmodified pavement-grade 64-22 asphalt binders upon cooling from 150 $^{\circ}$C to room temperature (RT) at 5 $^{\circ}$C/min, and changes in these microstructures when the surface was terminated with a transparent solid (glass) or liquid (glycerol) over-layer. The main findings are: (1) At free binder surfaces, wrinkled microstructures started to form near the wax crystallization temperature ($\sim$45 $^{\circ}$C), then grew to $\sim$5 $\mu$m diameter, $\sim$25 nm wrinkle amplitude and 10-30$\%$ surface area coverage upon cooling to RT, where they persisted indefinitely without observable change in shape or density.  (2) Glycerol coverage of the binder surface during cooling reduced wrinkled area and wrinkle amplitude three-fold compared to free binder surfaces upon initial cooling to RT; continued glycerol coverage at RT eliminated most surface microstructures within $\sim$4 hours.   (3) No surface microstructures were observed to form at binder surfaces covered with glass. (4) Sub-micron bulk microstructures were observed by near-infrared microscopy beneath the surfaces of all binder samples, with size, shape and density independent of surface coverage.  No tendency of such structures to float to the top or sink to the bottom of mm-thick samples was observed.  (5) We attribute the dependence of surface wrinkling on surface coverage to variation in interface tension, based on a thin-film continuum mechanics model. 



\end{abstract}

\maketitle


\section{\label{Introduction}Introduction}

Asphalt pavement is composed of a binder, generally bitumen, and mineral aggregates. While the aggregates make up the majority of the pavement by weight, the asphalt binder determines the pavement's strength and durability. Bitumen, a crude oil distillate, and is a complex mixture of long-chain hydrocarbons, some of which are waxes \cite{allen2014,loeber1998, redelius2015, sakib2018}. Atomic force microscopy (AFM) reveals a rich array of internally-textured microstructures at typical asphalt binder surfaces \cite{loeber1996}, while near infrared (NIR) optical microscopy reveals smaller, rounder microstructures distributed throughout the bulk \cite{ramm2016, ramm2018}. Some researchers have proposed that these features distribute stress within the binder, allowing for relaxation that affects the binder's micromechanical properties \cite{hung2015}. Engineering these properties of binders helps to develop pavements that resist rutting, cracking and other catastrophic deformations.

AFM topological images \cite{loeber1996} of free binder surfaces typically feature micrometer sized patches with several parallel wrinkles in their centers. The alternating height bands in the AFM images resemble the abdomen of a bee, hence have come to be labeled "bee" microstructures. Three different phases were initially identified in the surface topography: catana (the wrinkle), peri (the patch surrounding the wrinkle), and para (the matrix surrounding the peri phase). Lyne \textit{et. al.} suggested that the catana and peri phases were components of the same material phase with differing topography, called the laminate phase \cite{lyne2013}. A hypothesis about how bees form at free binder surfaces has begun to emerge from recent research. According to this hypothesis, thin film wax islands segregate at the hot binder surface. As the binder cools, the wax crystallizes and becomes stiffer than the surrounding matrix, causing it to wrinkle as it approaches room temperature \cite{hung2015,demoraes2010,schmets2010, pauli2011,das2013,blom2018,lyne2013}.

Here, to test this hypothesis, we study bee morphology and formation kinetics as a function of two variables: 1) Time during and after cooling the binder from 150 $^{\circ}$C to RT. These observations help to relate bee formation temperature to wax crystallization temperature. Optical (as opposed to atomic force) microscopy is the method of choice here, because the binder is fluid at elevated temperature. 2) Binder interface termination with glycerol or glass. These overlayers add differing amounts of interfacial tension, variations in which elucidate the connection between interface microstructure and interface tension. Moreover, neither overlayer reacts with bitumen, based on Hansen solubility parameters. Thus they modulate mechanical tension without influencing binder chemistry \cite{hansen2007}. Here, too, optical microscopy significantly augments AFM through its ability to image microstructures at interfaces buried beneath transparent over-layers, here solid glass and liquid glycerol.

The results of these studies show that bees indeed begin to form at a temperature that is consistent with the wax crystallization temperature. Moreover, we find that adding interfacial tension via glycerol or glass overlayers suppresses bee wrinkle amplitudes, in a manner consistent with thin film mechanics theory. We also find that extended glycerol coverage shrinks the laminate patch surrounding the wrinkle, even long after the sample has come to room temperature. 


\section{\label{Methods}Experimental Procedure}

\subsection{\label{sample prep}Sample preparation}

We performed all measurements with Superpave Performance Grade 64-22 binder, a typical grade for pavement in warm climates \cite{mcgennis}. We used the binder as-received from a producer who supplies asphalt binder in the state of Texas, which is representative of bitumen as it leaves a distilling plant.  No polymers or other chemicals were added, nor were any aging treatments applied.

\subsubsection{\label{Binder-glass interfaces}Air- and glass-binder interfaces}

\begin{figure}
\label{slide_sandwich_schematic}
\centering
\includegraphics[width=0.95\columnwidth]{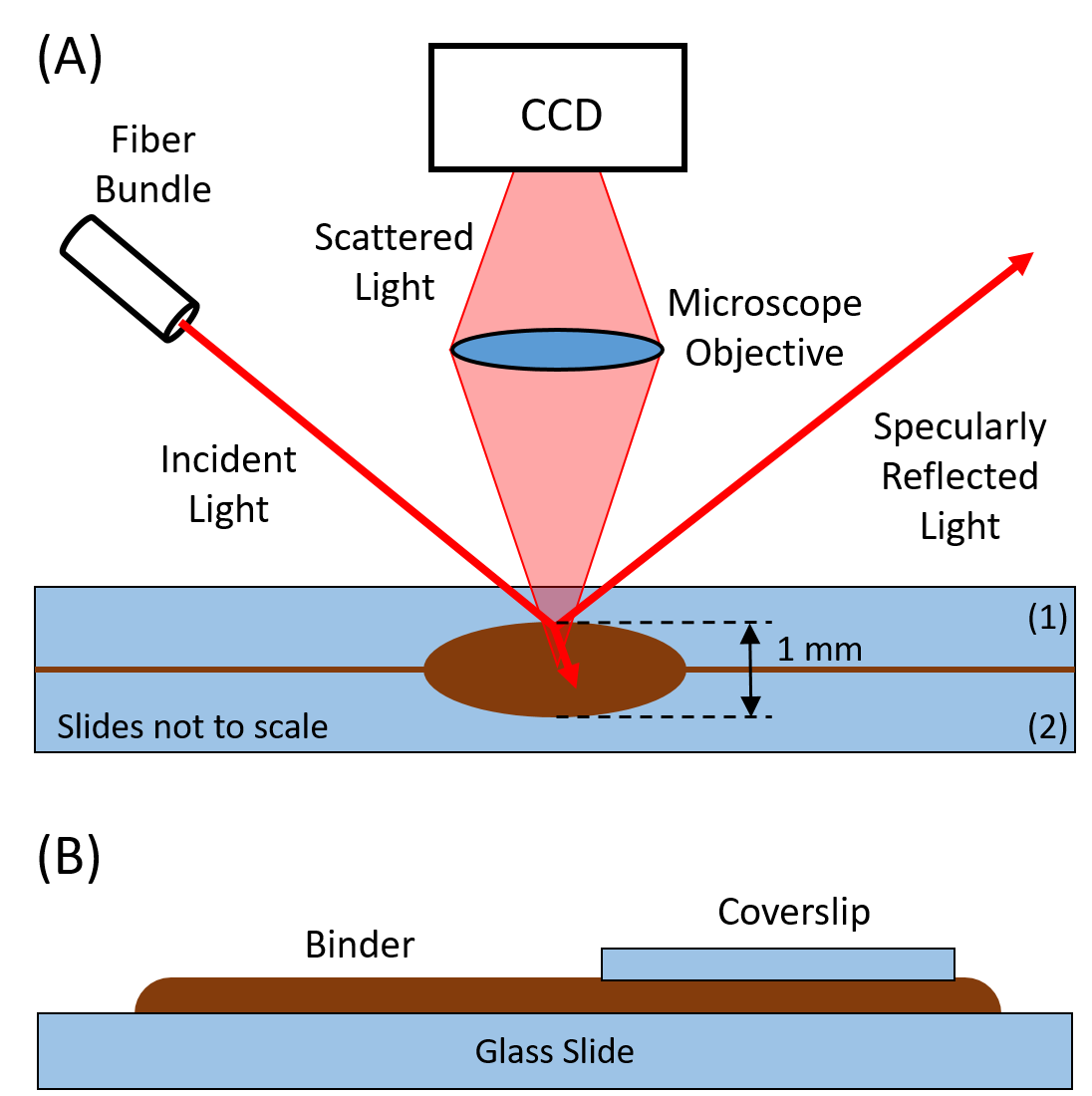}
\protect\caption{\label{fig:slide_sandwich_schematic} Schematic of slide-binder-slide sandwich and dark field optical microscope (A), and side by side air-binder and glass-binder interface sample (B).}
\end{figure}

To measure the temperature at which surface 'bee' microstructures first appeared at free surfaces, a 1-mm thick binder layer with area of $\sim$ 4 cm$^2$ was heated to 150 $^{\circ}$C on a flat microscope slide. The sample was then cooled at a rate of $\sim$5 $^{\circ}$C per minute while being imaged with the bright field optical microscope. Sample temperatures were recorded for each corresponding image.

To search for evidence that bulk microstructures, such as wax islands, float preferentially to the top interface of a sandwiched bitumen film, it was of interest to image both top and bottom glass-binder interfaces optically. We fabricated top-bottom-symmetric samples consisting of asphalt binder sandwiched between two 1 mm thick glass slides, each containing 0.5 mm deep dimples. The binder and slides were heated to 150 $^{\circ}$C, a small bead of binder was poured into the dimple of slide 1, then slide 2 was placed on top. The resulting sandwiched bitumen film was 1 mm thick in the area of observation (see Figure \ref{fig:slide_sandwich_schematic}A), thick enough that incident NIR light at could not penetrate through the sample and scatter off of the sample mount. Each sample sandwich was held at 150 $^\circ$C for 12 hours to ensure that any microstructure kinetics had plenty of time to reach a steady state. The sample was then cooled at 5 $^\circ$C/min to RT and both sides of the sample were imaged. We repeated each cycle with top and bottom slides inverted. 

To compare air-binder and glass-binder interfaces side by side, a 1 mm thick binder layer with area of $\sim$4 cm$^2$ was heated to 150 $^{\circ}$C on a flat microscope slide. A glass cover slip was then placed over half of the binder (Fig. \ref{fig:slide_sandwich_schematic}B) and the sample cooled to RT as for other samples. Air-binder and glass-binder interfaces were then imaged. Samples with exposed binder (Fig. \ref{fig:slide_sandwich_schematic}B) could not be inverted and imaged at elevated temperatures because the liquid binder flowed.




\subsubsection{\label{binder-air}Glycerol-binder interfaces}

\begin{figure}
\label{washer_schematic}
\centering
\includegraphics[width=.95\columnwidth]{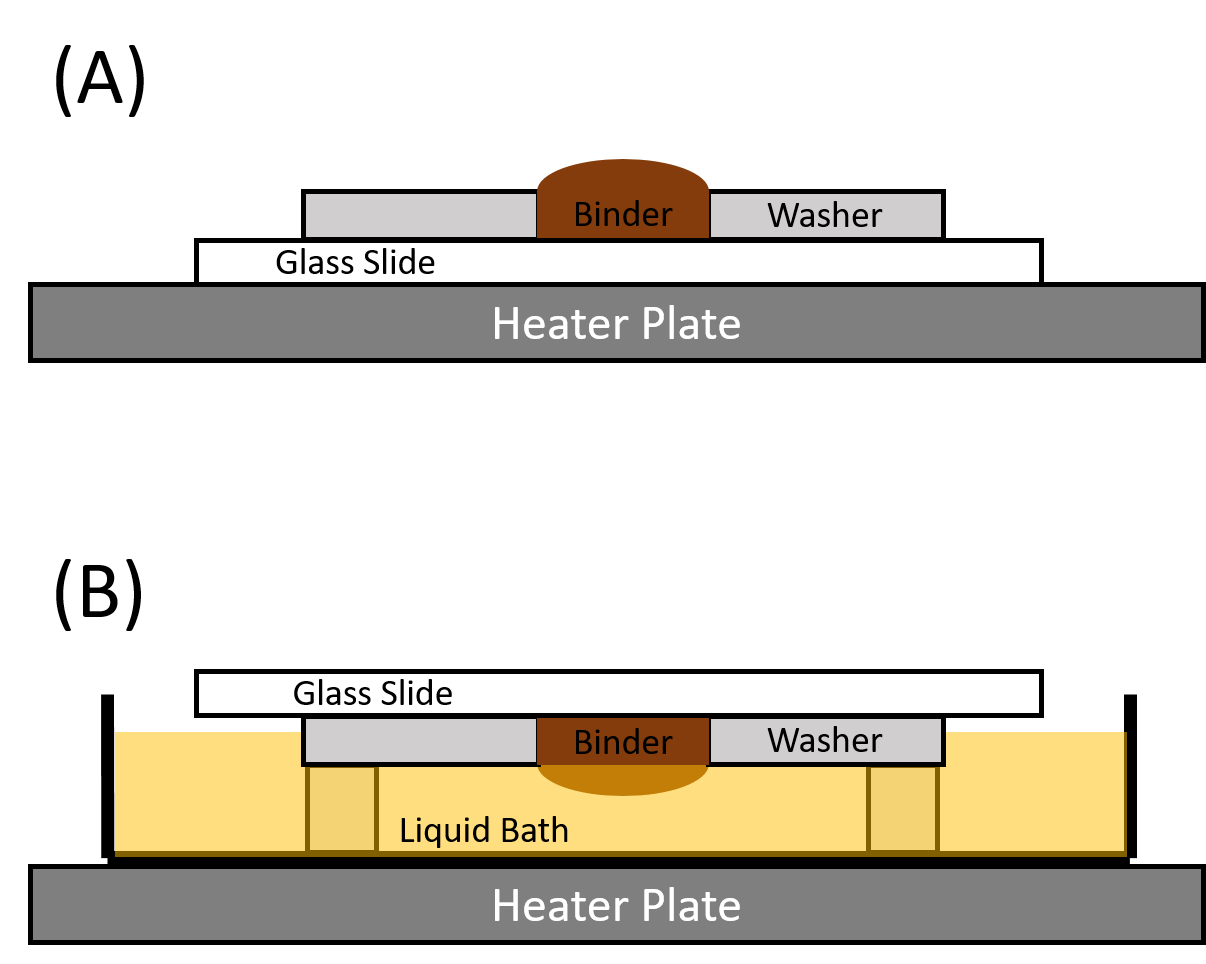}
\protect\caption{\label{fig:washer_schematic} Side view schematic of glycerol-binder interface setup. The sample is first prepared with hot binder poured into a washer on a glass slide, then cooled to room temperature (A). After reaching room temperature, the washer-slide-binder sample is placed upside down in a glycerol bath and heated to 150 $^{\circ}$C. Slice view through center of washer (B)}
\end{figure}

To heat and cool the asphalt binder with a liquid-binder interface, a washer supported and partially submerged in a tray of glycerol was used (see Figure \ref{fig:washer_schematic}). The sample was first prepared by heating a glass slide with steel washer on it to 150 $^{\circ}$C. Then 150 $^{\circ}$C binder was poured into the middle of the washer, filling the hole to a height above the top of the washer. The sample was then cooled to room temperature where the binder solidified and became one piece with the washer (see Fig. \ref{fig:washer_schematic}A).

\begin{figure}
\label{before_glycerol_after_bees}
\centering
\includegraphics[width=1.0\columnwidth]{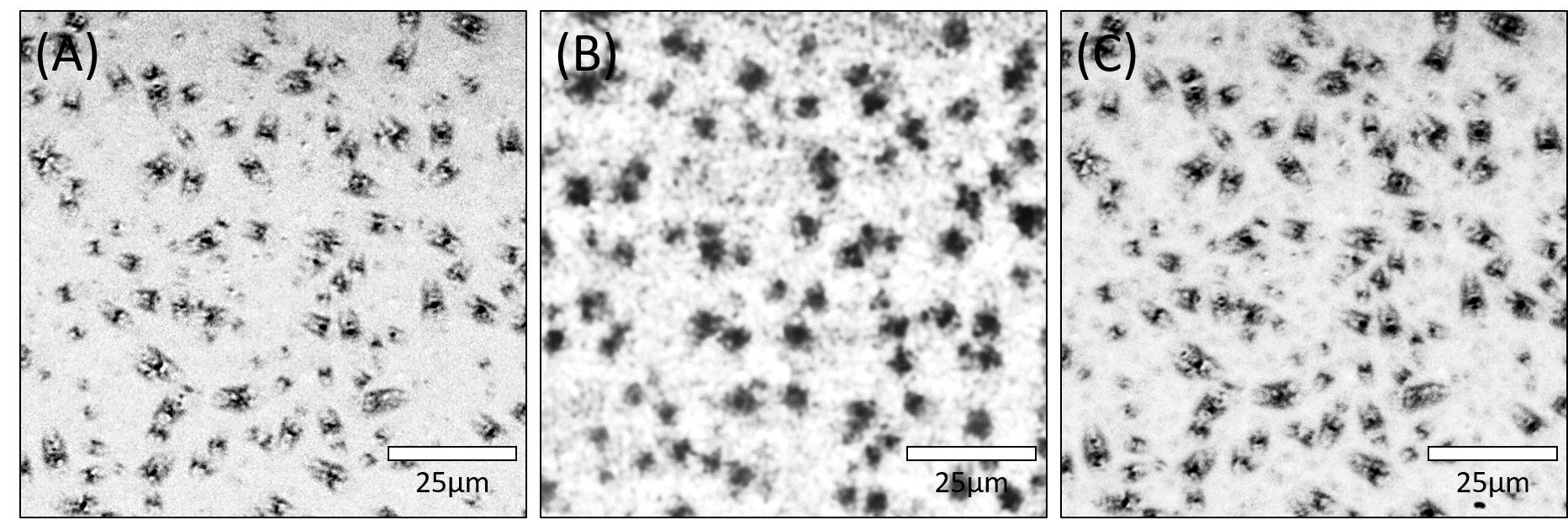}
\protect\caption{\label{fig:before_glycerol_after_bees} 550 nm reflected bright field optical microscope image sequence of free surface (A), glycerol covered (B), and post-rinsing (C) bees, all at RT.  Panels A and C show the same surface region, panel B a different, but similarly structured, region.}
\end{figure}

The sample was then inverted and placed in a glycerol bath (see Fig. \ref{fig:washer_schematic}B). The entire bath system was conditioned at 150 $^{\circ}$C for 15 minutes, and then cooled to RT at 5 $^\circ$C/min. The binder has a specific gravity less than glycerol, so it floated on top of the glycerol with a slightly convex surface, as seen in Figure \ref{fig:washer_schematic}B. Once cooled, the sample was removed from the glycerol and rinsed with water. We verified that this rinsing process did not affect the surface microstructure by recording the sequence of optical microscope images shown in Fig. \ref{fig:before_glycerol_after_bees}.  Fig. \ref{fig:before_glycerol_after_bees}A shows an image of bees at a RT air-binder interface, Fig. \ref{fig:before_glycerol_after_bees}B after covering the structures with glycerol, and Fig. \ref{fig:before_glycerol_after_bees}C, the same region as in Fig. \ref{fig:before_glycerol_after_bees}A, after rinsing with water.  Pre- and post-rinsing structures are indistinguishable. Rinsed samples were imaged with both an optical microscope and AFM as described in Sec. \ref{diagnosis}.


To observe the effect of prolonged glycerol coverage, the glycerol was left on the sample after cooling to room temperature for a designated amount of time (15, 30, 45, 60, and 240 minutes), then rinsed. Samples were then imaged as described above.

Since initial conditioning temperature and cooling rate influence bee formation \cite{merce2015,nahar2013}, control samples with an air-binder interface were fabricated with the same thermal history as the glycerol-binder interface samples. Specifically, one sample in each experimental set was placed in the glycerol bath with the binder right side up and above the surface of the glycerol, then subjected to the same thermal treatment as submerged samples.

\subsection{\label{diagnosis}Sample diagnosis}

\subsubsection{\label{dark field}Optical microscopy}

Figure \ref{fig:slide_sandwich_schematic}A depicts the dark field optical microscope. A halogen lamp illuminated the sample surface through a fiber bundle and interchangeable band-pass filters at 45 degree incidence angle. Most of the incident light was either specularly reflected or absorbed by the binder. A microscope objective (Nikon, 40X LWD) collected the small portion that surface and bulk microstructures scattered into a cone around the surface normal and imaged it onto a CMOS camera (Mightex). Dark field images were obtained at 500 $\pm$ 20 nm and 850 $\pm$ 50 nm incident wavelengths, for which bitumen penetration depths are $\sim$2$\mu$m and $\sim$25$\mu$m, respectively. Images at these wavelengths thus highlight primarily interface (500 nm) and bulk (850 nm) microstructures, respectively \cite{ramm2016}.

Reflected bright field optical microscope images were obtained with a Leitz Ergolux compound optical microscope equipped with 20X (Leitz NPL, 0.45 numerical aperature) and 100X (Olympus LMPLFLN, 0.80 numerical aperature, 3.4 mm working distance) objectives. A 20W halogen lamp with a 550 nm narrow band filter illuminated the sample. Images of surface structures were recorded with a 14 MP Amscope CMOS digital camera (Amscope MU1403). 

Both dark- and brght-field images were analyzed using the commercial software ImageJ \cite{imagej}. This analysis yielded the number of scattering centers per $\mu$m$^2$, and the average scatter intensity over a 2500 $\mu$m$^2$ area.

\subsubsection{\label{AFM}Atomic force microscopy}

AFM measurements were all made in dynamic contact (tapping) mode \cite{howland1996} on an Asylum Research MFP-3D Infinity AFM. In this configuration, the tip oscillates near its resonant frequency, "tapping" on the sample's surface. This mode measures surface height variations, as well as phase variations related to the sample's stiffness, adhesion and other material properties. For height measurements, a piezoelectric feedback system holds the tip at a set distance above the surface. As surface height changes, the cantilever reacts to changing forces (Van der Walls, electrostatic, dipole-dipole, etc.) and the feedback loop adjusts to the change. Simultaneously, the phase of cantilever oscillations varies in response to changing tip-surface interactions.

\section{\label{results}Results}

\subsection{\label{half covered samples dark field}Air- and glass-binder interface}

\begin{figure}
\label{bee_formation_sequence}
\centering
\includegraphics[width=1.0\columnwidth]{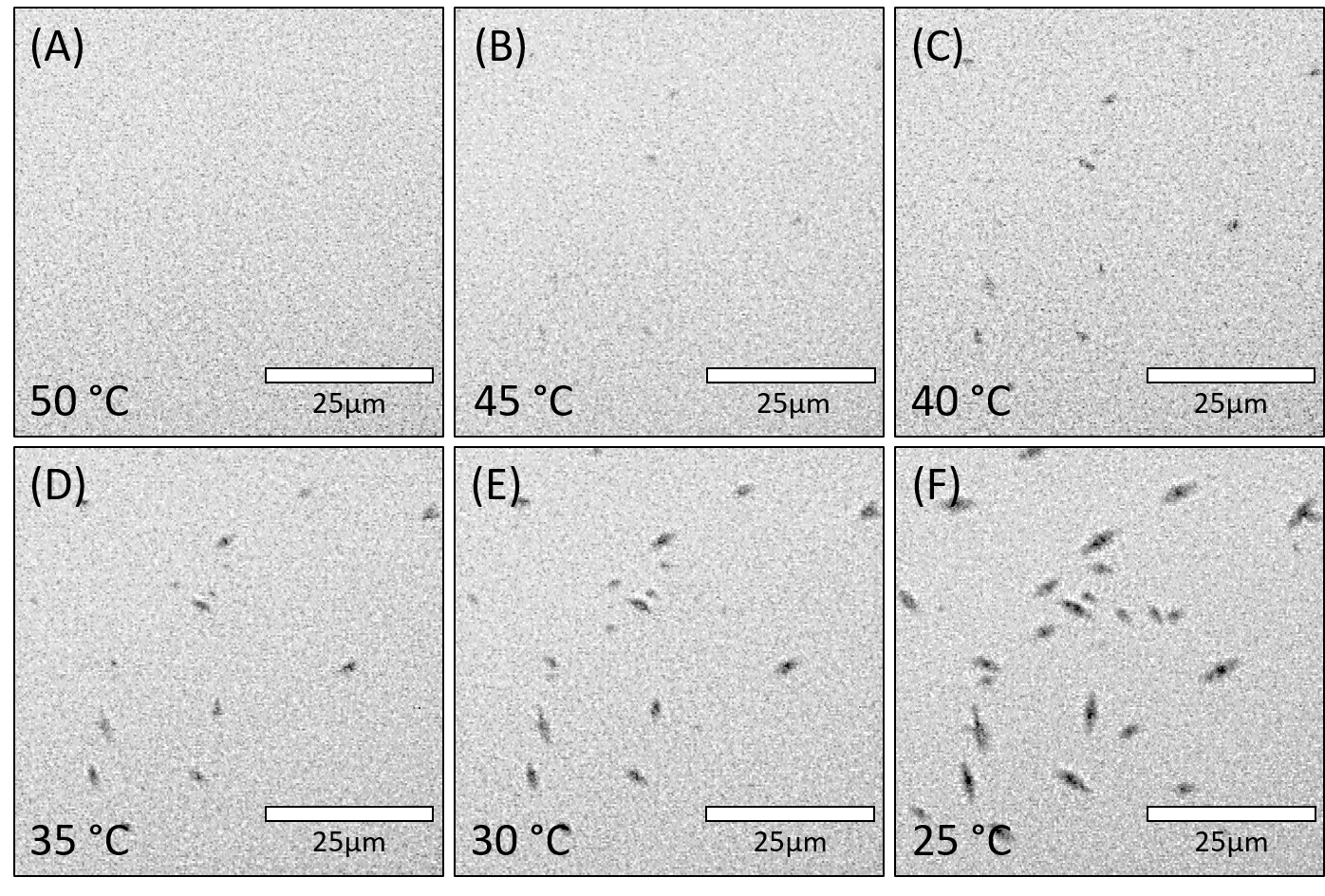}
\caption{\label{fig:bee_formation_sequence} Bright field optical microscope images of binder surface during cooling from 150 $^{\circ}$C to RT. Microstructures had not appeared at 50 $^{\circ}$C (A). Around 45 $^{\circ}$C microstructures began to appear (B). During further cooling the microstructured continued to grow and form into 'bee' microstructures (C-F). }
\end{figure}

\begin{figure}
\label{bees_covered_and_uncovered_2}
\centering
\includegraphics[width=.90\columnwidth]{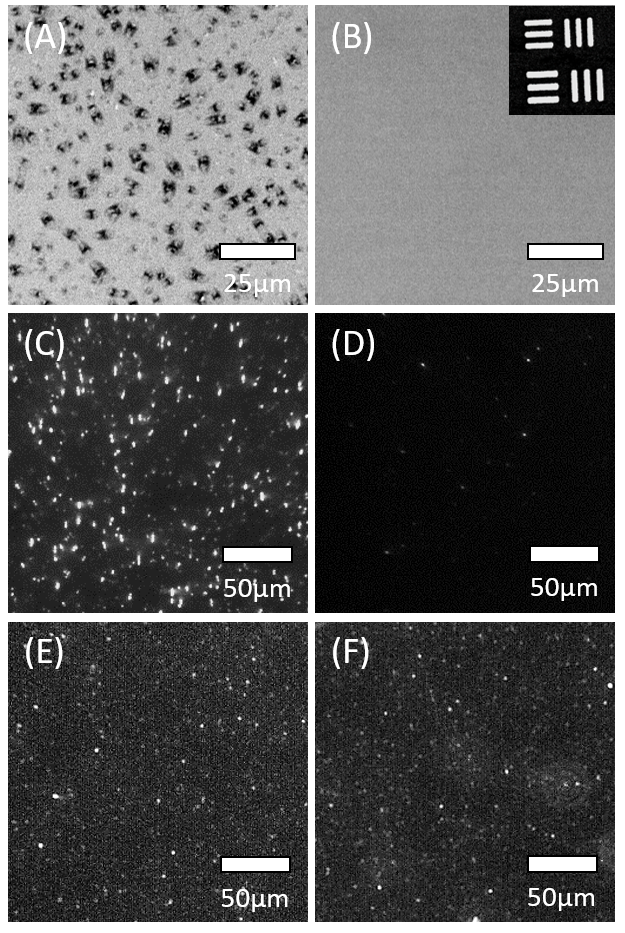}
\caption{\label{fig:bees_covered_and_uncovered_2} Optical microscope images of air-binder (left column:  A, C, E) and glass-binder (right column: B, D, F) interfaces at RT.  Bright- (A, B) and dark- (C,D) field images at 500 nm; dark-field images at 850 nm (E, F). Inset in image B shows resolution capability for glass covered samples.}
\end{figure}

\begin{table}
\centering
\setlength{\tabcolsep}{1em}
\renewcommand{\arraystretch}{1.5}
\label{tab_bee_formation}
\begin{tabular}{|c|c|}
\hline
Sample       & Bee Formation Temp. ($^{\circ}$C) \\ \hline
Unmodifed PG 64-22 & 48 $\pm$ 2 \\ \hline
1\% Sasobit PG 64-22 & 73 $\pm$ 1      \\ \hline

\end{tabular}
\caption{\label{tab_bee_formation} Bee formation temperatures for unmodified PG 64-22 binder and 1\% Sasobit modified PG 64-22 binder. Stated values are averages (formation temperatures) and standard deviations (uncertainties) obtained from nine (unmodified) or three (Sasobit-modified) specimens of each binder.}
\end{table}

Figure \ref{fig:bee_formation_sequence} shows a sequence of images of an air-binder interface as it cooled from 150 $^{\circ}$C to RT. At 50 $^{\circ}$C (Fig. \ref{fig:bee_formation_sequence}A), no microstructures were yet visible. At $\sim$45 $^{\circ}$C (Fig. \ref{fig:bee_formation_sequence}B), 1-2 $\mu$m diameter microstructures first became visible. Upon further cooling, the microstructures grow in size and number, transforming into wrinkled "bees" as the sample approaches RT (Figs. \ref{fig:bee_formation_sequence}C-F). Table \ref{tab_bee_formation} shows the temperatures at which the first surface microstructures were observed for unmodified PG 64-22 and 1\% wax additive (Sasobit) modified PG 64-22 (Sasol Performance Chemicals).

Figure \ref{fig:bees_covered_and_uncovered_2} shows bright (A, B) and dark (C-F) field optical microscope images for a sample with both an air-binder (left column: A, C, E) and a glass-binder (right column: B, D, F) interface. Bright field images at 550 nm show distinct bee microstructures at the air-binder interface (Fig. \ref{fig:bees_covered_and_uncovered_2}A), but none at the glass-binder interface (Fig. \ref{fig:bees_covered_and_uncovered_2}B). The inset in Figure \ref{fig:bees_covered_and_uncovered_2}B shows a USAF resolution test target imaged under a glass slide. The sharpness of this image shows that the microscope would have easily resolved 'bee'-sized features under the glass, had they been present.

Figures \ref{fig:bees_covered_and_uncovered_2}C and D show corresponding 500 nm dark field images. Bee microstructures at the air-binder interface scatter green light strongly and form bright scattering centers in the image. The glass-binder image, in contrast, shows hardly any such scattering centers, consistent with corresponding bright field image B. The same areas of the samples, imaged at 850 nm are shown in Figures \ref{fig:bees_covered_and_uncovered_2}E and \ref{fig:bees_covered_and_uncovered_2}F. Deeper penetration of near-IR light results in images that are similar to each other, dominated by smaller bulk scattering centers that do not depend on whether air or glass covers the film.


\begin{figure}
\label{dark_field_sandwich_unmod_3}
\centering
\includegraphics[width=0.9\columnwidth]{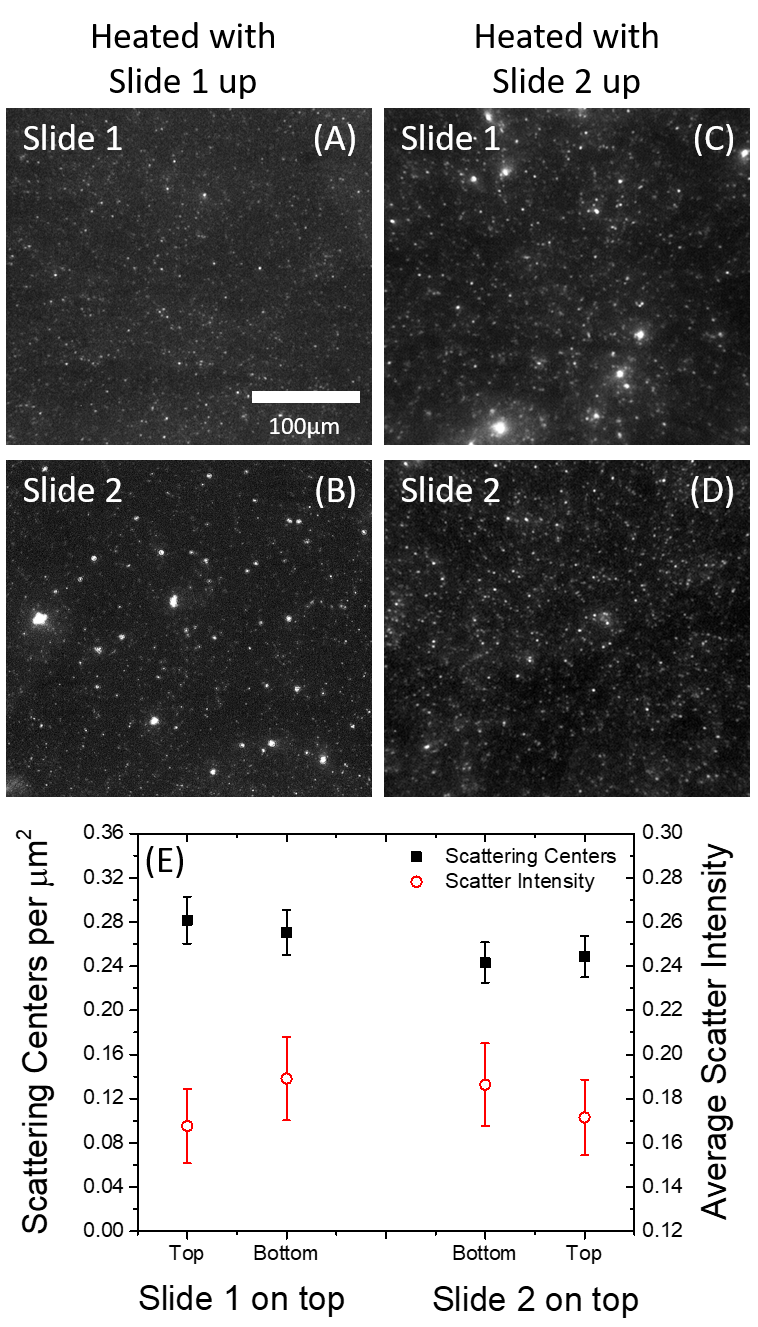}
\protect\caption{\label{fig:dark_field_sandwich_unmod_3} 850 nm dark field micrographs of top-bottom symmetric slide-binder-slide sandwich depicted in Fig. \ref{fig:slide_sandwich_schematic}A. A, B: Scattered light images from 25 $\mu$m bitumen layer beneath slide 1 (A) and slide 2 (B) when slide 1 is on top during conditioning at 150 $^\circ$C for 12 hours. C, D: Scattered light images for the same sample when inverted during conditioning. E: Scattering microstructure count, normalized per $\mu$m$^2$ (solid black squares) and normalized scatter intensity, averaged over 2500 $\mu$m$^2$ (open red circles). Error bars represent standard deviation over multiple areas of the same sample as well as multiple samples.}
\end{figure}

Figure \ref{fig:dark_field_sandwich_unmod_3} shows 850 nm optical dark field images of top (A, D) and bottom (B, C) glass-binder interfaces of the sandwich structure shown in Figure \ref{fig:slide_sandwich_schematic}A. As shown in Figures \ref{fig:bees_covered_and_uncovered_2}E and \ref{fig:bees_covered_and_uncovered_2}F, images at this wavelength depict bulk microstructures to a depth of $\sim$25 $\mu$m beneath each interface.


Figure \ref{fig:dark_field_sandwich_unmod_3}E plots the average scattering center counts per $\mu$m$^{2}$ and average scatter intensity over a 2500 $\mu$m$^{2}$ area. The left two sets of data points correspond to the first heating cycle where slide 1 was on top during heating and cooling. The right two sets of data points correspond to the second heating cycle where slide 2 was on top. Error bars represent a standard deviation of image measurements taken at multiple areas of each sample, and across multiple samples. Within these error bars, we observed no reproducible top-bottom asymmetry, nor any significant change with an inverted sample.

\subsection{\label{interfacial tension}Glycerol-binder interface}

\subsubsection{\label{initial cooling}Glycerol coverage during cooling}

\begin{figure}
\label{air_vs_glycerol_100X_and_AFM}
\centering
\includegraphics[width=1.0\columnwidth]{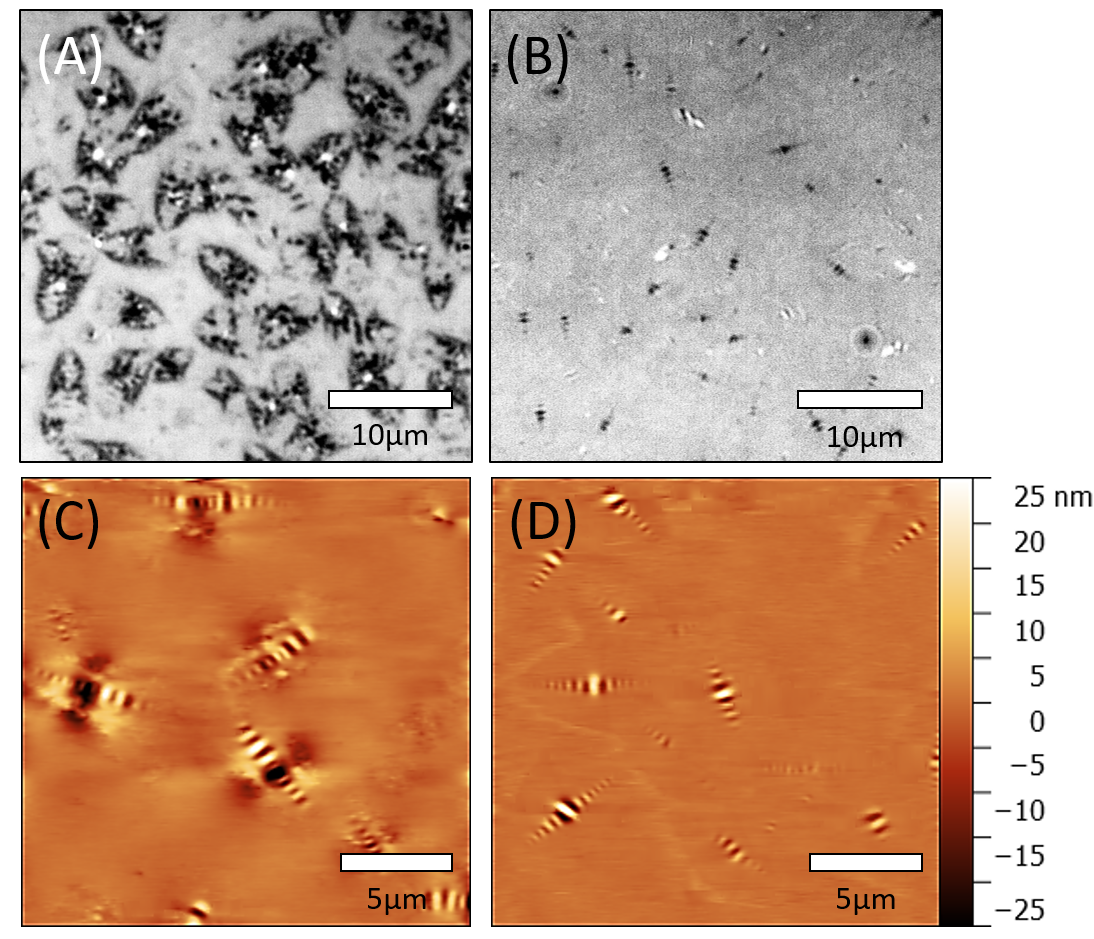}
\protect\caption{\label{fig:air_vs_glycerol_100X_and_AFM} Optical 550 nm bright-field (top row: A, B) and AFM height (bottom row: C, D) images of surface bee structures at air-binder (left column: A, C) and glycerol-binder (right column: B, D) interfaces, after cooling to RT, immediately rinsing, then waiting 24 hours.  Color scale for C, D indicates height above surface. Each microscope and AFM image pair is from a different location on the same sample surface.}
\end{figure}

\begin{table}
\centering
\setlength{\tabcolsep}{1em}
\renewcommand{\arraystretch}{1.5}
\label{tab_bee_amp}
\begin{tabular}{|c|c|c|}
\hline
Sample       & \multicolumn{2}{c|}{Wrinkle Amplitude (nm)} \\ \hline
             & Air-binder   & Glycerol-binder \\ \hline
1 & 24 $\pm$ 5.9 nm  & 11 $\pm$ 5.1 nm      \\ \hline
2 & 27 $\pm$ 6.1 nm  & 6.2 $\pm$ 4.3 nm      \\ \hline
3 & 15 $\pm$ 5.8 nm  & 5.1 $\pm$ 3.0 nm      \\ \hline
\end{tabular}
\caption{\label{tab_bee_amplitude} Average bee wrinkle amplitude for air-binder and glycerol-binder interfaces, measured by AFM for three different binders. Uncertainties represent the standard deviation of 15 to 30 bee measurements for each sample.}
\end{table}


Figure \ref{fig:air_vs_glycerol_100X_and_AFM} shows optical bright-field (top row: A, B) and AFM height (bottom row: C, D)  images of binder cooled to RT with an air-binder (left column: A, C) and a glycerol-binder (right column: B, D) interface. Glycerol was rinsed from the latter sample immediately upon reaching RT. Both samples were then held for 24 hours in air at RT before imaging. In optical images, the entire peri-catana structure of  air-interface bees is visible (Fig. \ref{fig:air_vs_glycerol_100X_and_AFM}A), while only the center wrinkle (catana) is visible at the glycerol-interface (Fig. \ref{fig:air_vs_glycerol_100X_and_AFM}B).

Table \ref{tab_bee_amplitude} shows the average amplitude of the highest central \cite{dossantos2014} wrinkle of 5-10 representative bees, obtained from AFM height images of each of three samples. Bee wrinkle amplitudes for glycerol-interface samples 1, 2 and 3 were 54\%, 77\%, and 66\% lower than for air-interface amplitudes, respectively.

\subsubsection{\label{post-cooling}Continued glycerol coverage after cooling}

\begin{figure}
\label{24_hours_glycerol_afer_cool_2}
\centering
\includegraphics[width=1.0\columnwidth]{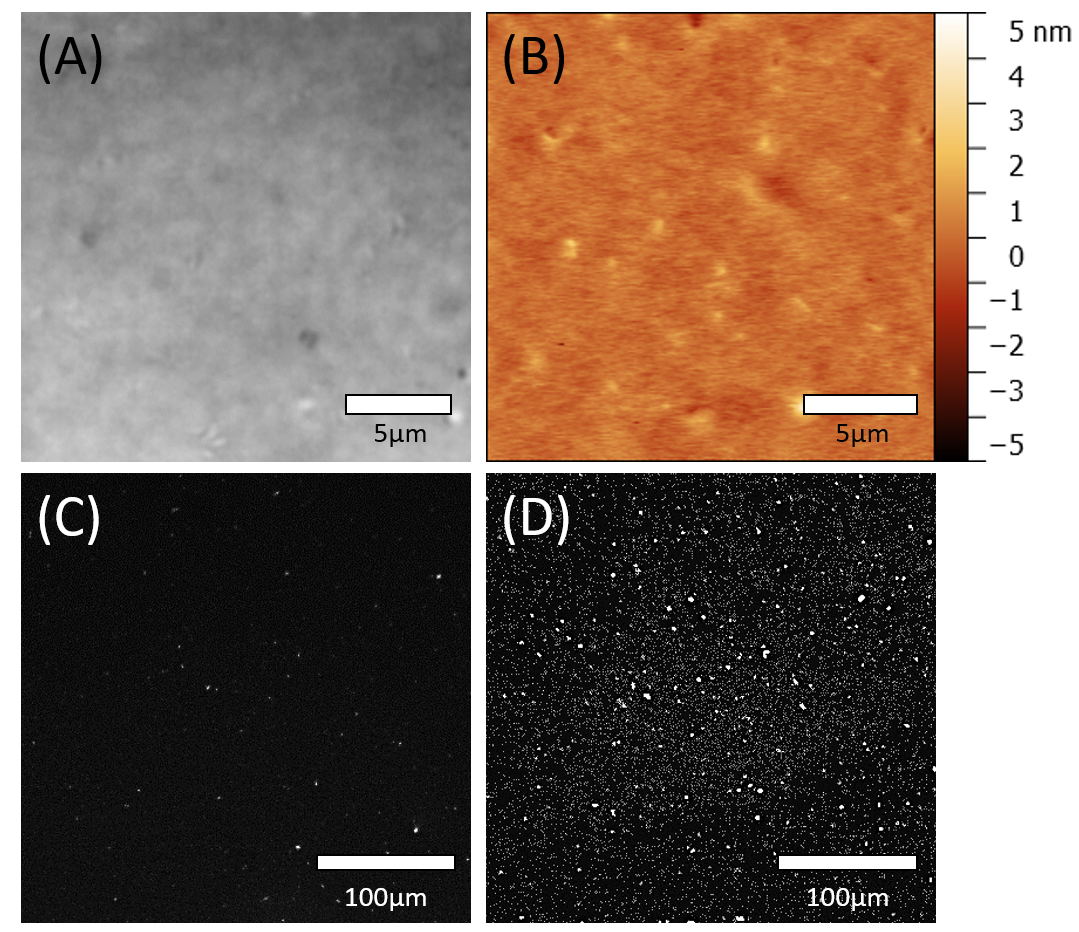}
\protect\caption{\label{fig:24_hours_glycerol_afer_cool_2} Optical bright-field micrograph (A) and AFM height image (B) of glycerol-binder interface cooled to RT and rinsed 24 hours later. Color scale in (B) shows height above surface. Optical dark field image at 500 nm (C), confirming lack of surface microstructure as in (A) and (B); dark-field image at 850 nm (D), showing persistence of bulk microstructure.}
\end{figure}

\begin{figure*}
\label{6_liquid_RT_times_with_phase}
\centering
\includegraphics[width=2.0\columnwidth]{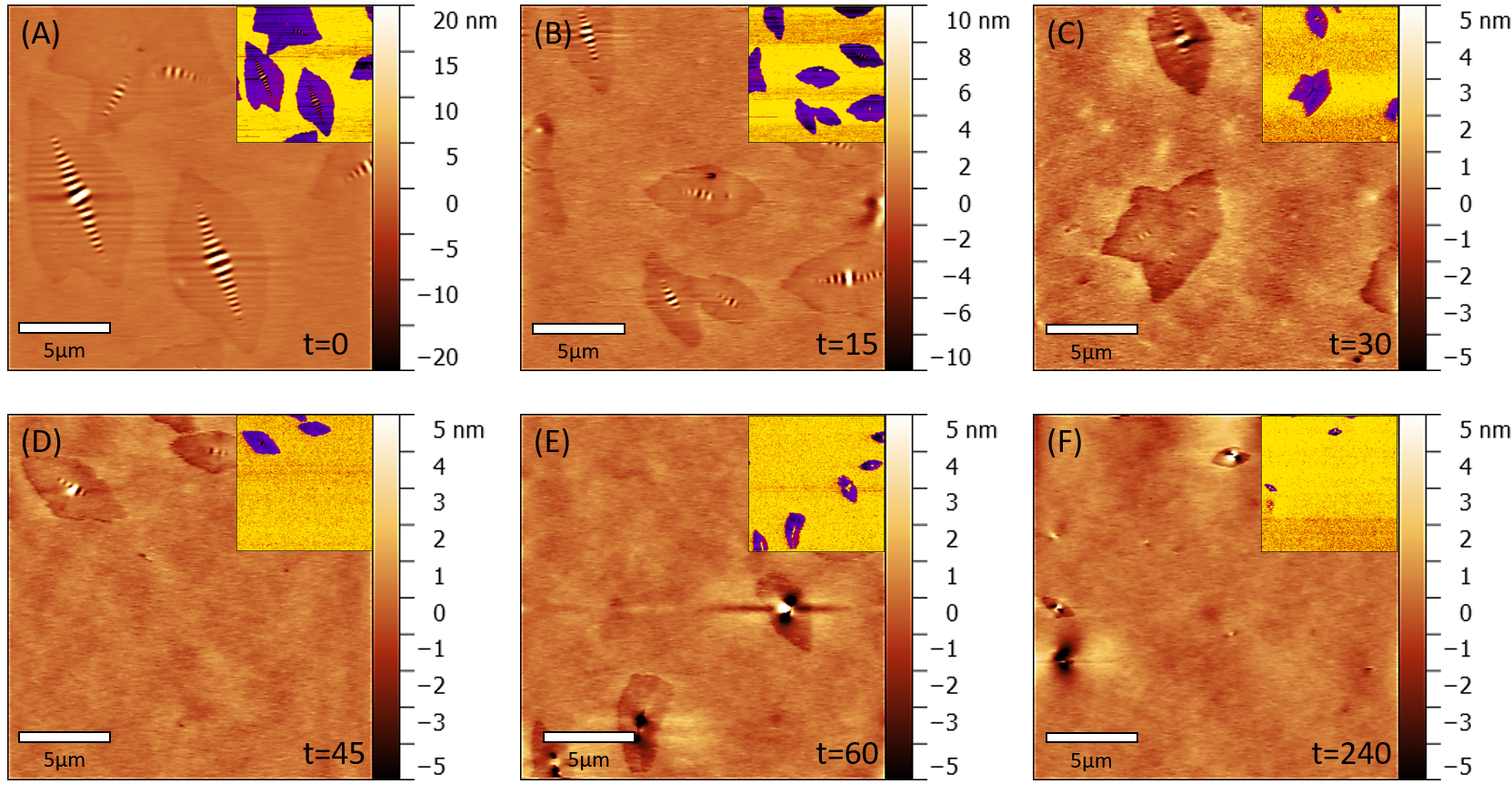}
\protect\caption{\label{fig:6_liquid_RT_times_with_phase} AFM height images of binder conditioned in glycerol for (A) 0, (B) 15, (C) 30, (D) 45, (E) 60, and (F) 240 minutes after cooling to RT. Samples were then rinsed, left in ambient air for 24 hours, and imaged. Insets show AFM phase contrast images.}
\end{figure*}

Figure \ref{fig:24_hours_glycerol_afer_cool_2} shows optical bright-field (A), AFM (B), and optical dark field (C, D) microscope images for glycerol-binder interface samples that were left in the glycerol bath for 24 hours after cooling to RT, rather than rinsing them immediately as for results in Fig. \ref{fig:air_vs_glycerol_100X_and_AFM}. Prolonged RT exposure to glycerol has now erased bee structures (see Fig. \ref{fig:24_hours_glycerol_afer_cool_2}A and \ref{fig:24_hours_glycerol_afer_cool_2}B). The AFM phase images (not pictured) also showed no differentiation of phase anywhere on the sample, as is seen when bees are present. Figures \ref{fig:24_hours_glycerol_afer_cool_2}C and \ref{fig:24_hours_glycerol_afer_cool_2}D show dark field optical microscope images at 500 nm and 850 nm, respectively. The 500 nm image confirms the absence of surface scattering centers, while the 850 nm image shows that the smaller bulk scattering centers persist, as in Figs. \ref{fig:bees_covered_and_uncovered_2}E and \ref{fig:bees_covered_and_uncovered_2}F.

Figure \ref{fig:6_liquid_RT_times_with_phase} shows AFM height (main panels) and corresponding phase (insets) images of rinsed binder surfaces following RT glycerol coverage for times ranging from 0 (A) to 4 hours (F). These images show progressively lower bee wrinkle amplitudes, as well as shrinking laminate phase size and areal density (see image insets) with increasing RT glycerol coverage time. Substantial bee suppression is evident even within the first hour (Panels A-E).

\section{\label{discussion}Discussion}

Bee formation temperatures reported in Fig. \ref{fig:bee_formation_sequence} and Table \ref{tab_bee_formation} can be compared with wax crystallization temperatures of various binders determined by differential scanning calorimetry (DSC) \cite{soenen2014,Lu2007}. Soenen \textit{et al.} report wax crystallization occurring at $\sim$43 $^\circ$C in unmodified binders and at $\sim$73 $^\circ$C in Sasobit modified binder \cite{soenen2014}. These agree very well with bee formation temperatures that we observed for similar binders as shown in Table \ref{tab_bee_formation}. This coincidence supports the hypothesis that "bee" microstructures grow from crystalline wax films that nucleate at the surface as it cools to the wax crystallization temperature \cite{hung2015}.  Further cooling prompts this film to grow, while simultaneously subjecting it to thermal contraction mismatch with the cooling substrate.  The resulting compressive strain causes the wax film to wrinkle, creating its characteristic "bee"-shaped structure \cite{hung2015}.  


The wax to form these surface islands must originate from the underlying bulk. The ubiquitous bulk microstructures that IR dark-field microscopy reveals immediately beneath a binder surface (see e.g. Figs. \ref{fig:bees_covered_and_uncovered_2}E,F and \ref{fig:24_hours_glycerol_afer_cool_2}D) also form near the wax crystallization temperature \cite{ramm2016, ramm2018}, and may therefore represent bulk counterparts of surface wax islands.  Nevertheless, the results in Fig. \ref{fig:bees_covered_and_uncovered_2} revealed no preferential concentration of such structures immediately beneath the top, or below the bottom, interface of a binder film, as might be expected if differences in specific gravity propelled them either to float to the top or sink to the bottom.  It is therefore likely that surface/interface tension, rather than gravity, is the primary driver of wax segregation at the binder surface.  The strong influence of glass (Fig. \ref{fig:bees_covered_and_uncovered_2}) and glycerol (Figs.  \ref{fig:air_vs_glycerol_100X_and_AFM}-\ref{fig:6_liquid_RT_times_with_phase}) on "bee" formation corroborates this conclusion.  Such over-layers alter surface/interface tension, but do not affect specific gravity of local wax concentrations.



The partial or total suppression of wrinkle amplitude by glycerol (Table \ref{tab_bee_amplitude}) or glass (Fig. \ref{fig:bees_covered_and_uncovered_2}) over-layers, respectively, also points to the dominant role of interfacial tension in governing "bee" formation. A basic model of wrinkle formation based on strain energy, as described by Huang (2005) \cite{huang2005}, predicts the observed trend qualitatively.

For an elastic film on a thick substrate, total wrinkle strain energy per unit area $U_{Total}$ can be resolved into three contributions
\begin{equation}\label{U_total}
    U_{Total} = U_{Bending} + U_{Compression} + U_{Substrate},
\end{equation}

where $U_{Bending}$ is the in-plane bending strain energy, $U_{Compression}$ is the in-plane compression strain energy, and $U_{Substrate}$ is the strain energy from the substrate. The last term can be ignored here since the substrate is much thicker than the film. $U_{Total}$ is a function of wavenumber $k$ and amplitude $A$.

A liquid over-layer adds a new strain energy term $U_{Tension}$ which can be modeled as the interfacial tension multiplied by the wrinkle length per wavelength
\begin{equation}\label{U_tension_1}
    U_{Tension} = \frac{1}{\lambda} L \gamma,
\end{equation}

where $\lambda$ is the wrinkle wavelength, $L$ is the distance along the wrinkle's oscillating surface, and $\gamma$ is the interfacial tension between the thin film and the liquid over-layer. If we model the wrinkle as a cosine function
\begin{equation}\label{wrinkle}
    w = A \cos{(k x)},
\end{equation}

with amplitude $A$ and wavenumber $k$, then $L$ is the arc length of $w$ for one period
\begin{equation}\label{arc_length}
    L = \int_{0}^{\lambda} \sqrt{1 + \left(\frac{dw}{dx}\right)^2} dx
\end{equation}

Combining Eqs. \ref{U_tension_1}, \ref{wrinkle} and \ref{arc_length}, the interfacial tension strain energy density becomes
\begin{equation}\label{U_tension_2}
    U_{Tension} = \frac{1}{\lambda} \left[ \int_{0}^{\lambda} \sqrt{1 + k^{2} A^{2} \left(sin(kx)\right)^2} dx \right] \gamma.
\end{equation}

Typical values for $\lambda$ and $A$ are around 500 nm and 25 nm, respectively. From this we can calculate that $(kA)^2 \sim 0.1$, and Eq. \ref{U_tension_2} can be Taylor expanded and integrated to solve for $U_{Tension}$ as a function of interfacial tension, wavenumber and amplitude
\begin{equation}\label{U_tension_3}
    U_{Tension} \approx \gamma + \frac{1}{4} k^2 A^2 \gamma.
\end{equation}

 Huang (2005) has shown that the other contributions to $U_{Total}$ are also functions of $k$ and $A$ \cite{huang2005}. Elastic thin film wrinkle theory shows that the wrinkle wavelength depends solely on the thickness of the film and on the elastic moduli of the film and substrate \cite{huang2006}, none of which vary when interfacial tension increases as a result of e.g. adding an over-layer. Therefore when interfacial tension is added, the amplitude A is the only parameter that is free to vary. The addition of $U_{Tension}$ increases total strain energy. In order to minimize the strain energy in the system, the wrinkle amplitude $A$ will decrease, consistent with observations presented in Figure \ref{fig:air_vs_glycerol_100X_and_AFM} and Table \ref{tab_bee_amplitude}.

In a purely elastic system, the duration of glycerol coverage after the sample had reached room temperature would not affect bee formation. However, bitumen is viscoelastic \cite{poel1954}, so time is also a variable in bee formation and suppression. The delayed suppression of bees seen in Figures \ref{fig:24_hours_glycerol_afer_cool_2} and \ref{fig:6_liquid_RT_times_with_phase} reflects the binder's viscoelasticity.  Indeed, the $\sim$1 hour time constant for bee suppression shown in Fig. \ref{fig:6_liquid_RT_times_with_phase} is similar to the time constant for viscoelastic creep in rheological and microstructural properties of PG 64-22 binder following a temperature increment \cite{ramm2018}.

\section{\label{conclusion}Conclusion}

This study used optical and atomic-force microscopy to elucidate mechanisms by which wrinkled microstructures ("bees") form at surfaces of pavement grade asphalt binders as they cool from 150 $^{\circ}$C to RT.  Real-time visible-wavelength (500 nm) optical microscopy, which is selectively sensitive to surface structures, showed that bees form at the wax crystallization temperature, determined independently by calorimetry.  Microscopy at more deeply penetrating near-infrared wavelengths (850 nm) showed that underlying bulk microstructures remain isotropically distributed throughout the bulk, with no observable tendency to float or sink despite 12 hours of conditioning at 150 $^{\circ}$C.  Combined optical and atomic-force microscopy measurements showed that adding interfacial tension to the cooling binder by covering its surface with glycerol significantly decreases the size, wrinkle amplitude and number density of the surface "bees".  Covering the binder's surface with glass suppresses bee formation altogether.  Yet neither over-layer had an observable influence on underlying bulk microstructure. 

Taken together, results from this study support the hypothesis that bee microstructures at free surfaces of asphalt binders are thin films of wax extruded from the underlying bulk by surface tension that grow and wrinkle upon cooling below the wax crystallization temperature.  They also show that surface "bees" differ fundamentally from underlying bulk microstructure not only in their larger size and intricate internal morphology, as shown previously \cite{ramm2016, ramm2018}, but in their sensitivity to surface conditions, especially surface tension.  Nevertheless, both types of structures form at similar temperatures, suggesting that they may share a similar chemical make-up.

\section{\label{acknowledgments}Acknowledgments}

The authors acknowledge support of Robert Welch Foundation grant F-1038 and National Science Foundation grant CMMI-1053925.

\bibliographystyle{ieeetr}
\bibliography{bib2}

\end{document}